\documentclass[twocolumn,twoside,superscriptaddress,showpacs,prd,10pt,aps]{revtex4-1}

\usepackage{graphicx}
\usepackage{amsmath}
\usepackage{hyperref}
\usepackage{color}

\RequirePackage{xspace}

\begin{document}

\title{Effect of direct $CP$ violation in charm on $\gamma$ extraction 
       from $B^{\pm}\to DK^{\pm}$, $D\to K^0_S\pi^+\pi^-$ Dalitz plot analysis}

\author{Alex Bondar}
\affiliation{Budker Institute of Nuclear Physics SB RAS, Lavrentieva 11, Novosibirsk, 630090, Russia}
\affiliation{Novosibirsk State University, Pirogova 2, Novosibirsk, 630090, Russia}
\author{Alexander Dolgov}
\affiliation{Novosibirsk State University, Pirogova 2, Novosibirsk, 630090, Russia}
\affiliation{ITEP, Bol. Cheremushkinskaya ul., 25, Moscow, 113259, Russia}
\affiliation{Dipartimento di Fisica e Scienze della Terra, Universita degli Studi di Ferrara,
Polo Scientifico e Tecnologico --- Edificio C, via Saragat 1, 44122 Ferrara, Italy}
\affiliation{Istituto Nazionale di Fisica Nucleare, Sezione di Ferrara,
Polo Scientifico e Tecnologico --- Edificio C, via Saragat 1, 44122 Ferrara, Italy}
\author{Anton Poluektov}
\affiliation{Budker Institute of Nuclear Physics SB RAS, Lavrentieva 11, Novosibirsk, 630090, Russia}
\affiliation{Department of Physics, University of Warwick, Coventry CV4 7AL, United Kingdom}
\author{Vitaly Vorobiev}
\affiliation{Budker Institute of Nuclear Physics SB RAS, Lavrentieva 11, Novosibirsk, 630090, Russia}
\affiliation{Novosibirsk State University, Pirogova 2, Novosibirsk, 630090, Russia}

\date{\today}

\newcommand{\bdk}{\ensuremath{B^{\pm}\to DK^{\pm}}\xspace}
\newcommand{\bdpi}{\ensuremath{B^{\pm}\to D\pi^{\pm}}\xspace}
\newcommand{\dsdpi}{\ensuremath{D^{*\pm}\to D\pi^{\pm}}\xspace}
\newcommand{\dnkpp}{\ensuremath{\overline{D}{}^0\to K^0_S\pi^+\pi^-}\xspace}
\newcommand{\dkpp}{\ensuremath{D\to K^0_S\pi^+\pi^-}\xspace}
\newcommand{\dn}{\ensuremath{D^0}\xspace}
\newcommand{\dnbar}{\ensuremath{\overline{D}{}^0}\xspace}
\newcommand{\kspp}{\ensuremath{K^0_S\pi^+\pi^-}\xspace}
\newcommand{\ad}{\ensuremath{\mathcal{A}{}_D}\xspace}
\newcommand{\adbar}{\ensuremath{\overline{\mathcal{A}}{}_D}\xspace}

\newcommand{\new}[1]{\textcolor{red}{#1}}

\begin{abstract}
  A possible effect of direct $CP$ violation in \dkpp decay on the $\gamma$ measurement 
  from \bdk, \dkpp Dalitz plot analysis is considered. Systematic uncertainty of $\gamma$
  coming from the current limits on direct $CP$ violation in \dkpp is estimated, 
  and a modified model-independent procedure of \bdk, \dkpp Dalitz plot analysis 
  is proposed that gives an unbiased $\gamma$ measurement even in presence of direct $CP$
  violation in charm decays. The technique is applicable to other three-body $D$ decays
  such as $D^0\to K_S^0K^+K^-$, $D^0\to \pi^+\pi^-\pi^0$, {\it etc.}
\end{abstract}


\maketitle

\section{Introduction}

The mechanism of $CP$ violation in particle physics is of primary importance because 
of its impact on cosmological baryogenesis and possible antimatter existence in the universe. 
In the quark sector, $CP$ violation is studied by measuring the elements of Cabibbo-Kobayashi-Maskawa (CKM)
mixing matrix~\cite{Cabibbo:1963yz, Kobayashi:1973fv} with the convenient representation given by  
the Unitarity Triangle (UT), the angles and sides of which are parameters observable in various decays of $B$ mesons. 
Precision measurement of the UT angle $\gamma$ (also denoted as $\phi_3$) is an essential 
ingredient in searches for New Physics phenomena in $B$ decays. The value of $\gamma$
acts as one of the Standard Model reference points against which other measurements of 
the UT parameters are compared. The irreducible theoretical uncertainty in the extraction 
of the angle $\gamma$ from \bdk decays is due to electroweak corrections and is extremely small, 
of the order of $10^{-6}$~\cite{Zupan:2011mn}. However, the experimental determination of $\gamma$ value 
remains a challenge owing to the low probabilities of the decays involved. 

The types of measurements that dominate $\gamma$ sensitivity are based on \bdk decays where the neutral $D$
meson decays into a $CP$ eigenstate (commonly referred to as the GLW method~\cite{Gronau:1990ra, Gronau:1991dp}), 
suppressed $K\pi$ state 
(ADS method~\cite{Atwood:1996ci}), or self-conjugate three-body final state such as \kspp
(GGSZ or Dalitz plot method~\cite{Giri:2003ty, Bondar:2002hd}). None of these methods are systematically  
limited at the current level of precision. However, obtaining a degree-level precision on $\gamma$
will require some subtle effects to be taken into account. The effect of charm 
mixing has already been considered by several authors~\cite{Silva:1999bd, Grossman:2005rp, Bondar:2010qs} 
and was found to be negligible in most cases. Recent evidence of direct $CP$ violation in singly Cabibbo-suppressed
two-body $D$ decays reported by LHCb~\cite{Aaij:2011in} has triggered discussions of the 
possible effect of $CP$ violation in charm on $\gamma$ measurements using 
the GLW technique~\cite{Wang:2012ie, Martone:2012nj, Bhattacharya:2013vc}. 
Although the evidence of large $CP$ violation in $D\to hh$ decays is not supported by the 
updated LHCb measurements~\cite{Aaij:2013bra, LHCb-CONF-2013-003}, this effect can play its role 
in precision measurements of $\gamma$. 

In this paper, we consider a possible effect of direct $CP$ violation in charm on the measurement 
of $\gamma$ using the Dalitz plot analysis of \bdk, \dkpp decays. 
The decay \dkpp is dominated by Cabibbo-favored transitions, and thus direct $CP$ asymmetry 
coming from the Standard Model effects is expected to be very small. However, future precision 
measurements of $\gamma$ can reach the point where this contribution will become significant. 
On the other hand, if disagreement in the UT parameters due to New Physics will be found in future measurements, 
the method that can distinguish whether the New Physics contribution enters charm or $B$ decays will be essential. 
In addition, ``effective'' $CP$ violation in \dkpp decay of the order of $10^{-3}$ should arise from the 
$CP$ violation in the neutral kaon system if this effect it not explicitly accounted for. 

The goal of this paper is twofold. First, we 
estimate the systematic uncertainty on $\gamma$ coming from the current limits on direct $CP$ violation in the 
\dkpp decay. Second, we show that in the model-independent analysis using quantum-correlated $D\overline{D}$
data at charm threshold it is possible to account for the $CP$ violation in charm and obtain an unbiased 
measurement of $\gamma$ without significantly sacrificing the statistical precision. 

Although the decay \dkpp is used throughout this paper, the same approach can be applied to other three-body 
final states of the neutral $D$ decay, such as $D^0\to K_S^0K^+K^-$, $D^0\to \pi^+\pi^-\pi^0$, {\it etc.}

\section{Formalism of model-independent $\gamma$ measurement with $CP$ violation in \dkpp}

\label{sec:formalism}

The procedure to extract $\gamma$ from \bdk, \dkpp in a model-independent way employed in the current 
analyses~\cite{Libby:2010nu, Aihara:2012aw, Aaij:2012hu} assumes $CP$ conservation in $D$ decays. 
The technique uses binned Dalitz plot distributions, and in order to utilize the assumption of 
$CP$ conservation the bins are chosen symmetrically to the exchange of Dalitz plot 
variables of the \dkpp decay (invariant masses squared of $K^0_S\pi^+$ and $K^0_S\pi^-$ combinations): 
$m^2_{K_S^0\pi^+}\leftrightarrow m^2_{K_S^0\pi^-}$. The bins are denoted 
with the index $i$ which runs from $-\mathcal{N}$ to $\mathcal{N}$ excluding zero; symmetric bins 
have the same $|i|$ and the flip of the sign $i\leftrightarrow -i$ corresponds to the reflection 
$m^2_{K_S^0\pi^+}\leftrightarrow m^2_{K_S^0\pi^-}$. Current analyses~\cite{Libby:2010nu, Aihara:2012aw, Aaij:2012hu} 
are performed with 
$\mathcal{N}=8$: this number of bins, together with a special choice of the shape of the bins 
over the Dalitz plot, provides a statistical precision for the $\gamma$ measurement that approaches 
the precision of the unbinned model-dependent technique with the limited \bdk data available 
today~\cite{Bondar:2008hh}. 

The procedure of model-independent Dalitz plot analysis is described in detail in 
Refs.~\cite{Giri:2003ty, Bondar:2008hh}; here we give only the final equations. 
The analyses use four categories of events involving \dkpp: flavor-tagged
\dkpp, neutral $D$ mesons tagged in $CP$ eigenstate from $\psi(3770)\to D\overline{D}$ process ($D_{CP}$), correlated 
pairs of neutral $D$ mesons where both $D$ are reconstructed in the \kspp state, and \bdk decays with \dkpp. 
The number of events in bin $i$ of the flavor-tagged $D$ decay is different for \dn and \dnbar; it is 
denoted as $K_i$ and $\overline{K}{}_i$, respectively. However, in the $CP$-conserving case and with the 
symmetric binning described above $K_{i} = \overline{K}{}_{-i}$, so $\overline{K}{}_i$ are not independent. 
The numbers of events in bins are then related as
\begin{equation}
  M_i = h_{CP} \left[K_i + K_{-i} + 2\sqrt{K_i K_{-i}} C_i\right]
  \label{n_cp}
\end{equation}
for $D$ decays into a $CP$ eigenstate, 
\begin{equation}
  \begin{split}
  M_{ij} = h_{\rm corr} & \left[ K_i K_{-j} + K_{-i}K_j - \right. \\
                               & \left. 2\sqrt{K_i K_{-i}K_j K_{-j}} (C_i C_j + S_i S_j)\right] \\
  \end{split}
  \label{n_corr}
\end{equation}
for correlated $D\overline{D}$ pairs both decaying into \kspp (here one has to deal with two 
correlated Dalitz plots and thus the number of events is described with two indices $i$ and $j$), and  
\begin{equation}
  N^{\pm}_i = h_{B^{\pm}} \left[K_{\pm i} + r_{\pm}^2 K_{\mp i} + 2\sqrt{K_i K_{-i}}(x_{\pm} C_i \pm y_{\pm} S_i)\right] \\
  \label{n_b}
\end{equation}
for \dkpp from \bdk. 
Here $x_{\pm}=r_B\cos(\delta_B\pm\gamma)$, $y_{\pm}=r_B\sin(\delta_B\pm\gamma)$, $r^2_{\pm}=x^2_{\pm} + y^2_{\pm}$. 
The free parameters $x_{\pm}$ and $y_{\pm}$ hold the information about the phase $\gamma$ and hadronic 
parameters in \bdk decay: the amplitude ratio $r_B$ and the strong phase difference $\delta_B$. 
The other free parameters are the normalization
factors $h_{CP}$, $h_{\rm corr}$ and $h_{B^{\pm}}$, and phase terms $C_i$ and $S_i$. The terms $C_i$ and $S_i$
describe the average sine and cosine of the strong phase difference between \dn and \dnbar amplitudes 
over the bin $i$. In the case of $CP$ conservation they satisfy $C_i=C_{-i}$, $S_{i} = -S_{-i}$. Thus, 
these parameters are independent only for $i>0$. The system of equations (\ref{n_cp}), (\ref{n_corr}), and 
(\ref{n_b}) is overconstrained and can be solved with the maximum likelihood fit to obtain $x_{\pm}$ and $y_{\pm}$
and, thus, the value of $\gamma$. 

Now we turn to the case when $CP$ is not conserved in the \dkpp decay. The relations between 
symmetric bins of the Dalitz plot do not hold anymore. We still use the notation for bin number 
$i=-\mathcal{N},\ldots -1,1,\ldots \mathcal{N}$, but now 
$\overline{K}{}_{i}\neq K_{-i}$, $C_{i}\neq C_{-i}$, and $S_{i}\neq -S_{-i}$ in general. 
In principle, now the binning is not required to be symmetric, although  in our 
studies we keep the same binning as in the $CP$-conserving case to 
allow for a direct comparison of the two approaches. 

The equations relating the numbers of events in bins of the \dkpp Dalitz plots are: 
\begin{equation}
  M_i = h_{CP} \left[K_i + \overline{K}{}_{i} + 2\sqrt{K_i\overline{K}{}_{i}} C_i\right]\,,
  \label{ncpv_cp}
\end{equation}
\begin{equation}
  \begin{split}
  M_{ij} = h_{\rm corr} & \left[ K_i\overline{K}{}_{j} + \overline{K}{}_{i}K_j - \right. \\
                               & \left. 2\sqrt{K_i\overline{K}{}_{i}K_j\overline{K}{}_{j}} (C_i C_j + S_i S_j)\right]\,, \\
  \end{split}
  \label{ncpv_corr}
\end{equation}
and 
\begin{equation}
  \begin{split}
  N^{+}_i = & h_{B^{+}} \left[K_i + r_+^2 \overline{K}{}_{i} + 2\sqrt{K_i\overline{K}{}_{i}}(x_+ C_i + y_+ S_i)\right] \,, \\
  N^{-}_i = & h_{B^{-}} \left[\overline{K}_i + r_-^2 K_{i} + 2\sqrt{\overline{K}_i K_{i}}(x_- C_i - y_- S_i)\right]\,. \\
  \end{split}
  \label{ncpv_b}
\end{equation}
Note that the number of phase terms $C_i$, $S_i$ is doubled compared to the $CP$-conserving case since their 
values for $i<0$ are now independent. The numbers of flavor-tagged events $K_i$ and $\overline{K}{}_{i}$ also
have to be obtained independently, but since the available samples of flavor-tagged $D$ decays are large, 
this should not limit the accuracy of the measurement. There are $4\mathcal{N}+8$ free parameters for 
$4\mathcal{N}^2+6\mathcal{N}$ equations (\ref{n_cp}), (\ref{n_corr}), and (\ref{n_b}), 
and thus the system of equations still remains solvable. As a result, arbitrarily large $CP$ violation in 
the \dkpp decay does not lead to a bias in the measurement of the $x,y$ parameters, and, hence, of the value 
of $\gamma$ when using this technique. We remind that there is a principal ambiguity in this measurement: 
it is not sensitive to the simultaneous change of sign of all $S_i$ which causes the signs of $y_{\pm}$ observables 
to flip. This ambiguity is resolved by the weak model assumption that the \dkpp amplitude is described 
with a sum of Breit-Wigner amplitudes~\cite{Bondar:2008hh}. 

The decays of a neutral $D$ in a $CP$ eigenstate into \kspp are obtained from the process $\psi(3770)\to D\overline{D}$, 
where the other (tagging) $D$ meson is reconstructed in the $CP$ eigenstate of the opposite parity. Therefore, 
if $CP$ is violated in the decay of the tagging $D$, Eq. (\ref{ncpv_cp}) would not be valid. 
This effect is expected to be larger for $CP$-even tags using Cabibbo-suppressed decays 
($D\to K^+ K^-, \pi^+\pi^-$) than for $CP$-odd tags which are mostly Cabibbo-favored (such as $D\to K_S^0\pi^0$). 
Without the $CP$-tagged $D$ decay, the remaining equations (\ref{ncpv_corr}) and (\ref{ncpv_b}), 
which do not include $D$ decays other than $\kspp$, have two additional ambiguities. 
One is an additional discrete ambiguity: the simultaneous change of sign of all $C_i$ followed by a flip of 
$x_{\pm}$ signs. The choice between the two solutions can, though, be made using Eq.~(\ref{ncpv_cp}) 
with the good assumption that $CP$ violation in the tagging $D$ decay is small.
The other, more important ambiguity is the rotation by the arbitrary phase $\delta\phi$: 
\begin{equation}
\begin{split}
  C'_i &= C_i\cos\delta\phi - S_i\sin\delta\phi, \\
  S'_i &= S_i\cos\delta\phi + C_i\sin\delta\phi, 
\end{split}
\end{equation}
with the simultaneous rotation of $\gamma$ by the same value $\delta\phi$~\footnote{We are grateful to EPJ referee 
for pointing this out.}. Thus, the single decay mode 
\dkpp cannot resolve the $CP$-violating phases originating from $B$ and $D$ decays and $D_{CP}$ decay 
has to serve as a reference. Any $CP$-violating phase in this decay directly translates into the uncertainty 
on the angle $\gamma$. Generally, the analysis using only $B\to DK$ and $\psi(3770)\to D\overline{D}$ decays 
can be influenced by the common $CP$ violating phase in charm which directly affects the $\gamma$ measurements
but is not observable otherwise. 

The $CP$ violating phase $\delta\phi$ can be independently controlled in the $B$ decay where the 
$D^0-\overline{D}{}^0$ admixture appears with known $CP$-violating phase other than $\gamma$. 
This is possible using the decay $B^0\to D\pi^0$, \dkpp. Neutral $D$ in this decay is a coherent admixture of 
$D^0$ and $\overline{D}{}^0$ states determined by the CKM phase $\beta$~\cite{Bondar:2005gk}. 
Using the binned approach, the decay time distributions for 
$B^0$ and $\overline{B}{}^0$ mesons are
\begin{equation}
  \begin{split}
  \frac{dN^{\overline{B}{}^0\to D\pi^0}_{i}(t)}{dt} & = e^{-\frac{|t|}{\tau}}\left[
         K_i\cos^2\frac{\Delta m t}{2} + \overline{K}_i\sin^2\frac{\Delta m t}{2}-\right. \\
               &\left.  \sqrt{K_i\overline{K}_i}(S_i\cos 2\beta + C_i\sin 2\beta)\sin \Delta m t \right], \\
  \frac{dN^{B^0\to D\pi^0}_{i}(t)}{dt} & = e^{-\frac{|t|}{\tau}}\left[
         \overline{K}_i\cos^2\frac{\Delta m t}{2} + K_i\sin^2\frac{\Delta m t}{2}+\right. \\
               &\left.  \sqrt{K_i\overline{K}_i}(S_i\cos 2\beta + C_i\sin 2\beta)\sin \Delta m t \right], \\
  \end{split}
\end{equation}
where $t$ is the difference $t=t_{\rm sig}-t_{\rm tag}$ between the $B$ decay time and the time at which it was tagged
to be $\overline{B}{}^0$ or $B^0$, $\tau$ is the average neutral $B$ lifetime, and $\Delta m$ is the mass difference 
of the two $B$ mass eigenstates. In the relations above, we neglected the Cabibbo-suppressed
contribution to $B^0\to D\pi^0$ which is of the order of $|V_{ub}V^*_{cd}/V_{cb}V^*_{ud}|\simeq 0.02$. It introduces 
additional parameters similar to $B\to DK$ case (amplitude ratio $r_{D\pi^0}$ and strong phase $\delta_{D\pi^0}$)
which, however, can be obtained from data in the time-dependent analysis~\cite{Bondar:2005gk}.

The $CP$ violating phase in the \dkpp decay would enter the 
difference between the angles $\beta$ observed in $B^0\to D\pi^0$ and $B^0\to J/\psi K^0_S$ decays. 
The uncertainty in $\gamma$ will then be limited by the theoretical uncertainties in $\beta$
extraction from these decays (mostly from $B^0\to J/\psi K^0_S$ since $B^0\to D\pi^0$ is tree-dominated), 
and by the experimental precision of $\beta$ measurement in $B^0\to D\pi^0$. 
The analyses performed by Belle~\cite{Krokovny:2006sv} and BaBar~\cite{Aubert:2007rp} suggest that 
the precision that can be obtained with the Belle II experiment with the integrated luminosity 50 ab$^{-1}$
can be around 2$^{\circ}$. Additional charmed $B$ decays sensitive to $\beta$, {\it e.g.} 
$B^0\to D\pi^+\pi^-$~\cite{Latham:2008zs}, can be used to improve this precision, not only with Belle II, 
but also with the LHCb experiment. 


The technique described above can be applied not only to \dkpp decays with $CP$ violation, but also 
to other non-self-conjugate final states, such as $D\to K^0_S K^-\pi^+$, $D\to K^+\pi^-\pi^0$, etc.

\section{Bias of $\gamma$ measurement due to $CP$ violation in \dkpp decay}

\label{sec:gamma_bias}

In this section, we investigate how the current limits on $CP$ violation in the decay \dkpp affect the 
model-independent measurement of $\gamma$ using the current approach in which $CP$ conservation is assumed 
in charm decays. 

The first study placing limits on direct $CP$ violation in \dkpp decay has been performed by 
the CLEO collaboration~\cite{Asner:2003uz} and has recently been improved by CDF~\cite{Aaltonen:2012nd}. 
Both measurements use isobar formalism to 
parametrize the \dkpp decay amplitude and 
assume that a $CP$ asymmetry may appear in any of the quasi two-body amplitudes. 
Specifically, the amplitude is represented as the sum of resonance components 
for both the $D^0$ and $\overline{D}{}^0$ decays: 
\begin{equation}
  \begin{split}
  \ad = & a_0e^{i\delta_0} + \sum\limits_{j}A^+_j\mathcal{M}_j\,, \\
  \adbar = & a_0e^{i\delta_0} + \sum\limits_{j}A^-_j\overline{\mathcal{M}}{}_j\,,
  \end{split}
\end{equation}
where \ad and \adbar are the amplitudes of \dn and \dnbar decays, respectively, 
$a_0$ and $\delta_0$ are the amplitude and the phase of the non-resonant component (assumed to be $CP$-conserving), 
$\mathcal{M}_j$ and $\overline{\mathcal{M}}{}_j$ are the quasi two-body resonant matrix elements (typically 
the relativistic Breit-Wigner amplitudes), and $A^{\pm}_{j}$ are the (complex) amplitudes of 
the resonant components. 
In the $CP$-violating case, $A^{+}_j\neq A^{-}_j$. In general, the presence of a $CP$-violating amplitude can result in 
the difference of both the magnitudes and the phases of the complex numbers $A^+_j$ and $A^-_j$. 
The parametrization adopted by CLEO and CDF is
\begin{equation}
  A^{\pm}_j = a_j e^{i(\delta_j\pm \phi_j)}(1\pm b_j/a_j)\,,
  \label{eq:cpv_param}
\end{equation}
where $a_j$ and $\delta_j$ are $CP$-averaged parameters, 
while $b_j/a_j$ and $\phi_j$ are small parameters related to the $CP$ violation. 

\begin{figure}
  \begin{center}
  \includegraphics[width=0.4\textwidth]{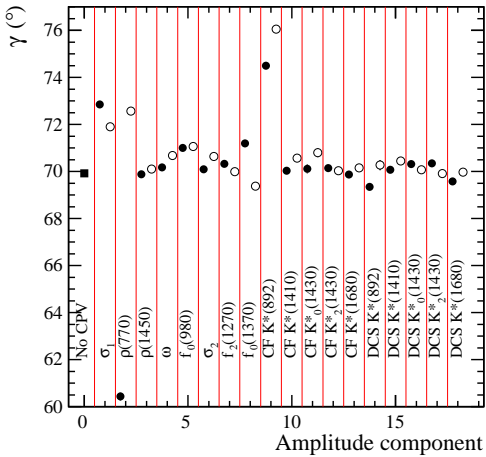}
  \end{center}
  \caption{Bias of $\gamma$ measurement in the fit without accounting for $CP$ asymmetry, 
           using the amplitude generated without $CP$ asymmetry (filled square), as well as
           with $CP$ asymmetry in the magnitude $b/a=10\%$ (filled circles), and in phase 
           $\phi=0.1$ (open circles) in each amplitude component. Statistical errors are 
           comparable with the size of markers. }
  \label{fig:bias_nocpv}
\end{figure}

We perform Monte Carlo (MC) studies to estimate the systematic uncertainty in the $\gamma$ measurement arising 
from the current limits on $CP$ violation in the \dkpp decay obtained by CDF. A parametrization of the form (\ref{eq:cpv_param})
is used. The \dkpp amplitude model used to generate event samples is based on the Belle measurement~\cite{Poluektov:2010wz}. 
A $CP$ asymmetry of 10\% is introduced one-by-one for the amplitude and phase in each partial amplitude ({\it i.e.}
$b_j/a_j=0.1$ and $\phi_j=0.1$ radian). Large number of flavor-tagged $D$, $D_{CP}$, correlated $D\overline{D}$, and \bdk
samples are generated so that the statistical error of $\gamma$ measurement does not exceed $0.2^{\circ}$.
The values $\gamma=70^{\circ}$, $r_B=0.1$, $\delta_B=130^{\circ}$ are used at the generation stage. 
The samples are then fitted to extract $\gamma$ value and other related parameters 
without taking $CP$ violation into account. The resulting values of $\gamma$ for each variation of the $D^0$ decay 
amplitude are shown in Fig.~\ref{fig:bias_nocpv}. 

After the bias in $\gamma$ for a 10\% $CP$ asymmetry is obtained from MC, we recalculate the 
bias associated with current experimental limits by taking the central values of $CP$ 
asymmetries and their errors from the CDF 
measurement~\cite{Aaltonen:2012nd} assuming a linear dependence of $\gamma$ bias on 
$CP$-violating parameters $b_j/a_j$ and $\phi_j$. The resulting contributions from $CP$ violation 
in each resonance are given in Table~\ref{tab:gamma_bias}. Finally, we calculate the 
total error by summing up the central values of bias linearly for each contribution, 
and the errors quadratically. The errors of the individual components are assumed to be 
uncorrelated. As expected, the resulting $\gamma$ bias is consistent with zero within its error 
since no evidence of $CP$ violation has been found by CDF. Thus the $\gamma$ 
uncertainty is taken as the error of the bias and amounts to around $3^{\circ}$. 

\begin{table}
  \caption{Contributions of $CP$ violating amplitudes in \dkpp decay 
           measured by CDF~\cite{Aaltonen:2012nd} to the $\gamma$ measurement bias for each 
           contributing resonance, and the total $\gamma$ bias. }
  \label{tab:gamma_bias}
  \begin{tabular}{|l|c|c|}
  \hline
 Resonance & \multicolumn{2}{c|}{Contribution to $\gamma$ bias (${}^{\circ}$)} \\
  \cline{2-3}
           & Amplitude & Phase \\
  \hline
       CF $K^{*}(892)$ & $+0.09 \pm 0.27$ & $-0.87 \pm 2.09$ \\
  CF $K^{*}_{0}(1430)$ & $-0.05 \pm 0.05$ & $-0.23 \pm 0.35$ \\
  CF $K^{*}_{2}(1430)$ & $+0.07 \pm 0.12$ & $-0.04 \pm 0.07$ \\
      CF $K^{*}(1410)$ & $+0.01 \pm 0.02$ & $-0.21 \pm 0.37$ \\
           $\rho(770)$ & $+0.27 \pm 0.89$ & $-0.24 \pm 0.97$ \\
              $\omega$ & $-0.32 \pm 0.21$ & $-0.25 \pm 0.36$ \\
          $f_{0}(980)$ & $-0.02 \pm 0.13$ & $-0.02 \pm 0.38$ \\
         $f_{2}(1270)$ & $-0.09 \pm 0.10$ & $-0.06 \pm 0.09$ \\
         $f_{0}(1370)$ & $-0.09 \pm 1.06$ & $+0.01 \pm 0.26$ \\
          $\rho(1450)$ & $-0.02 \pm 0.19$ & $-0.09 \pm 0.22$ \\
          $\sigma_{1}$ & $-0.31 \pm 0.78$ & $-0.09 \pm 0.62$ \\
          $\sigma_{2}$ & $-0.07 \pm 0.08$ & $-0.04 \pm 0.56$ \\
      DCS $K^{*}(892)$ & $-0.04 \pm 0.24$ & $+0.22 \pm 0.15$ \\
 DCS $K^{*}_{0}(1430)$ & $+0.23 \pm 0.44$ & $-0.12 \pm 0.21$ \\
 DCS $K^{*}_{2}(1430)$ & $-0.30 \pm 0.56$ & $+0.03 \pm 0.04$ \\
\hline
 Total & \multicolumn{2}{c|}{$-2.65 \pm 3.17$} \\
\hline
  \end{tabular}
\end{table}

\section{Model-independent analysis with $CP$ violation in \dkpp}

Here we present results of the MC study performed to estimate how the fit procedure which allows for $CP$
violation in \dkpp decay described in Section~\ref{sec:formalism} affects the statistical precision of the 
$\gamma$ measurement compared to the $CP$-conserving case. 

We have performed MC simulation with $10^{6}$ events of flavor-specific \dkpp of each flavor, 
120000 correlated $\psi(3770)\to D\overline{D}$ decays, 
120000 $D_{CP}\to K^0_S\pi^+\pi^-$ decays, and 
60000 \bdk, \dkpp decays of each $B$ sign. This $B$ sample size 
corresponds roughly to the data sample expected in the upgraded phase of the LHCb experiment and at the 
Super $B$ factory. The ratio of $\psi(3770)$ and $B$ samples was taken to be the same as in current analyses; 
we expect that the sufficient sample of $\psi(3770)$ decays will be collected by BES-III experiment and 
future tau-charm factory. 
In addition to this sample denoted by the factor $k=1$, we repeat the simulation with four times smaller ($k=1/4$) 
and four times larger ($k=4$) samples to check how the error scales with the sample size. 
Current world-average values for the parameters of $B\to DK$ decays are taken: 
$\gamma=70^{\circ}$, $r_B=0.1$, $\delta_B=130^{\circ}$. A total of 1000 pseudoexperiments are generated and 
fitted for the sample of each size. 

\begin{figure}
  \begin{center}
  \includegraphics[width=0.4\textwidth]{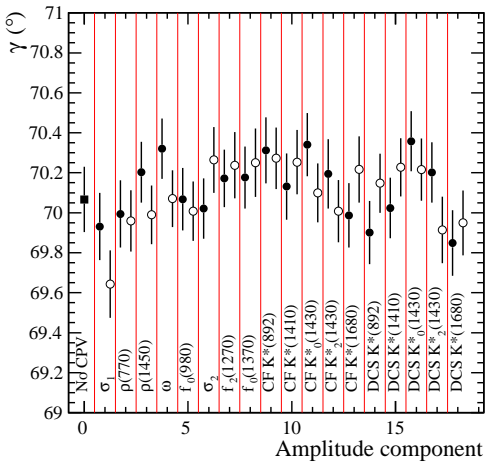}
  \end{center}
  \caption{Bias of $\gamma$ measurement in the fit with $CP$ asymmetry accounted for, 
           using the amplitude generated without $CP$ asymmetry (filled square), as well as
           with $CP$-violation in the magnitude $b/a=10\%$ (filled circles), and in phase 
           $\phi=0.1$ (open circles) in each amplitude component. Note different vertical 
           scale compared to Fig.~\ref{fig:bias_nocpv}}. 
  \label{fig:bias_cpv}
\end{figure}

Each MC sample is fitted with two techniques: a) the one which assumes $CP$ conservation in \dkpp decay, and 
b) the one which allows for $CP$ violation, as described in Section~\ref{sec:formalism}. Using the similar variations of the \dkpp 
amplitude involving 10\% $CP$ asymmetry in each resonance component as in Section~\ref{sec:gamma_bias}, 
we show that the bias in $\gamma$ measurement is consistent with zero (see Fig.~\ref{fig:bias_cpv}). 
We also compare the statistical precision of the two approaches from the spread of fitted $\gamma$ values 
between pseudoexperiments. As shown in Table~\ref{gamma_stat}, 
the reduction of statistical precision due to increased number of free parameters does not exceed 10\%. 

\begin{table}
  \caption{Comparison of the precision of the model-independent $\gamma$ measurement 
           for the fit procedures with and without accounting for the $CP$ violation 
           in \dkpp decay. The size of the simulated sample defined by the factor $k$
           is described in the text. }
  \label{gamma_stat}
  \begin{tabular}{|l|c|c|c|}
  \hline
  $k$ & $\sigma(\gamma)$ (No $CP$, ${}^{\circ}$) & $\sigma(\gamma)$ (with $CP$, ${}^{\circ}$) & Ratio \\
  \hline
  $1/4$  & $2.932\pm 0.081$ & $3.021\pm 0.084$ & $1.030\pm 0.040$ \\
  1  & $1.525\pm 0.042$ & $1.612\pm 0.049$ & $1.057\pm 0.043$ \\
  4 & $0.713\pm 0.019$ & $0.775\pm 0.019$ & $1.088\pm 0.039$ \\
  \hline
  \end{tabular}
\end{table}

\section{Conclusion}

We have shown that the current best limits on $CP$ violation in \dkpp decay coming from the measurement performed by 
CDF~\cite{Aaltonen:2012nd} translates to systematic uncertainty in the determination of the CKM phase 
$\gamma$ from \bdk, \dkpp decay of the order of $3^{\circ}$. While the current world-average precision of $\gamma$
is $9-12^{\circ}$~\cite{ckmfitter, utfit} and is not limited yet by this uncertainty, the data sample to be collected 
by LHCb experiment before its upgrade should allow measurement with a precision around $5^{\circ}$ 
in which the \bdk, \dkpp Dalitz analysis will have significant weight~\cite{Bediaga:2012py}. 
It is thus useful to study the  
$CP$ asymmetry in \dkpp with a larger data sample ({\it e.g.} at $B$ factories and LHCb) to reduce this 
uncertainty. 

In addition, we have shown that even if the \dkpp decay is found to exhibit $CP$ violation, it is possible 
to account for it and perform the unbiased measurement of $\gamma$ in \bdk, \dkpp decay in a 
model-independent way. Compared to the model-independent technique which assumes $CP$ conservation in 
\dkpp~\cite{Giri:2003ty, Bondar:2008hh}, this method has more free parameters which, however, leads to a
reduction of the statistical precision not exceeding 10\%. This approach reduces the possibly large 
number of $CP$-violating degrees of freedom in \dkpp amplitude to a single $CP$-violating phase which
directly affects the measurement of $\gamma$. This phase can be controlled using $D\overline{D}$
threshold data where the \dkpp decay is tagged by the other $D$ decaying to the $CP$-eigenstate through 
Cabibbo-favored transition ({\it e.g.} $K^0_S\pi^0$). However, this procedure should assume the absence 
of $CP$ violation in the tagging decay and thus is model-dependent. Another possibility 
is to access this phase from the difference of measurements of the angle $\beta$ in 
$B^0\to J/\psi K^0_S$ and $B^0\to D\pi^0$, \dkpp decays. The accuracy of this approach will be limited by 
the experimental precision of $\beta$ determination from $B^0\to D\pi^0$, \dkpp decays (about $2^{\circ}$ 
with Belle II), but can be improved further by using other modes with \dkpp, such as $B^0\to D^0\pi^+\pi^-$.

\begin{acknowledgments}
The authors are grateful to their colleagues from the LHCb collaboration for their interest to this study, 
and especially to Tim Gershon and Guy Wilkinson for useful comments and corrections. 
The work of A.B., A.D., and V.V. is partly supported by the grant of the Government of the 
Russian Federation (No 11.G34.31.0047). A.P. is supported by the Science and Technology 
Facilities Council (United Kingdom).
\end{acknowledgments}

\bibliography{modind_cpv}

\end{document}